\documentclass[12pt]{article}
\usepackage{graphicx,subeqn,epsf}
\def\be{\begin{equation}}
\def\ee{\end{equation}}
\def\bea{\begin{eqnarray}}
\def\eea{\end{eqnarray}}
\def \lsim{\mathrel{\vcenter
     {\hbox{$<$}\nointerlineskip\hbox{$\sim$}}}}

\def \delr{\mathrel{\vcenter
     {\hbox{$~\Delta$}\nointerlineskip\hbox{$\longrightarrow$}}}}
\def \rB{\mathrel{\vcenter
     {\hbox{$~{\kappa}$}\nointerlineskip\hbox{$\longrightarrow$}}}}

\begin{document}

\begin{flushright}
TIFR/TH/03-22
\end{flushright}

\vspace*{1cm}

\begin{center}
{\Large\bf History of Exotic Meson (4-quark) and \\ Baryon (5-quark)
States} \\[2cm] 
D.P. Roy 

\vspace*{.5cm}

Department of Theoretical Physics \\
Tata Institute of Fundamental Research \\ 
Homi Bhabha Road, Mumbai 400 005, INDIA
\end{center}

\vspace*{1cm}

\begin{center}
\underbar{\bf Abstract}
\end{center}
\bigskip

I briefly review the history of exotic meson (4-quark) and baryon
(5-quark) states, which is rooted in the formalism of Regge pole and
duality.  There are robust model-independent predictions for the
exchange of 4-quark (Baryonium) Regge trajectories in several
processes, which are strongly supported by experiment.  On the other hand
the predictions for the spectroscopy of 4-quark resonances are based
on specific QCD inspired models, with some experimental
support.  The corresponding predictions for the recently discovered
exotic baryon (Pentaquark) state are briefly discussed.

\newpage

\noindent {\Large\bf Introduction:}
\medskip

Recently several experiments have reported an exotic baryon resonance
$\Theta^+$ (1540) in the $K^+ n$ channel \cite{one}.  It has a narrow
width consistent with the experimental resolution $(\Gamma < 20 \ {\rm
MeV})$, and a tentative spin of ${1 \over 2}$.  Its exotic quantum
number, $Y = B + S = 2$, implies that it must belong to the
$\overline{10}$ or higher representation of the flavour $SU(3)$.  Its
minimal quark content is $(udud\bar s)$.  Indeed it represents the
first credible signal of an exotic baryon (Pentaquark) state.  At the
same time the BABAR, CLEO and the BELLE collaborations have found a
narrow charmed meson in the $D^+_s \pi^0$ channel at 2317 MeV
\cite{two}.  Again the width of this resonance is less than the
experimental resolution ($\Gamma \leq 10$ MeV), while the tentative
spin-parity assignment is $0^+$.  Although it has the quantum numbers
of a $c\bar s$ pair, the low mass, small width and the decay modes of
this resonance are very different from the potential model predictions
for a $0^+ c\bar s$ state.  Therefore it has been suggested as a
possible 4-quark $(cq\bar s\bar q)$ state.  Moreover BES colaboration
have also reported a narrow peak near the $\bar pp$ threshold as a
possible 4-quark state \cite{three}.  These experimental
discoveries have revived a great deal of interest in the long sought
for exotic meson $(qq\bar q\bar q)$ and baryon $(qqqq\bar q)$ states.

I shall outline here the history of the exotic meson and baryon
physics, discuss its main phenomenological achievements and
limitations, and finally look at its implications for the above
mentioned exotic baryon (Pentaquark) state.

Many of the present players in this field may not know that the
history of exotic meson and baryon states predates QCD.  It is based
in the $S$-Matrix approach to strong interaction, i.e. the formalism
of Regge pole and duality.  The Regge poles provide an effective
description of composite particles like mesons and baryons, where the
spin $J$ is strongly correlated with the particle mass $M$.  The main
meson and baryon states are seen to lie on linear trajectories in a
$J-M^2$ plot with a roughly universal slope, $\alpha'_{m,b} \simeq 1$
GeV$^{-2}$, and intersepts $\alpha_m (0) \simeq 0.5, \ \alpha_b (0)
\simeq 0$.  Thus the meson and baryon exchanges in the cross-channels
are described by virtual particle exchange amplitudes with effective
spins of $\alpha_m (t)$ and $\alpha_b (u)$.  The resulting energy
dependence of forward and backward meson-baryon scattering amplitudes
at high energy, $s^{\alpha_m (t \sim 0)}$ and $s^{\alpha_b (u \sim
0)}$, describe the high-energy soft scattering processes remarkably
well.  Moreover when extrapolated to lower energies, they provide a
good average description of the $s$-channel resonance contributions,
which shows a duality between the Regge pole and resonance
contributions.  Thus the Regge pole and duality formalism provides a
consistent and economical parametrisation of a host of soft two-body
scattering processes \cite{four}.  Subsequently this has been
extended to inclusive scattering cross-section, \be a + b \rightarrow
c + X,
\label{one}
\ee
which is related via optical theorem to a 3-body forward elastic scattering
amplitude $A_{ab\bar c}$ \cite{five}.  The Regge pole and duality
formalism for this amplitude provides an effective description of
inclusive scattering as well as resonance production in the channel
$X$ \cite{four,six}. 
\bigskip

\noindent {\Large\bf Prediction of Exotic Meson $(qq\bar q\bar q)$ and
Baryon $(qqqq\bar q)$ States:}
\medskip

It was shown in \cite{seven,eight} that the meson-meson and meson-baryon
scattering processes can be consistently described in the framework of
Regge pole and duality using only meson states in the 1,8 and baryon
states in the 1,8,10 representations of the flavour $SU(3)$; but a
consistent description of baryon-antibaryon scattering requires the
presence of exotic meson states in the representations
$10,\overline{10}$ and $27$.  Consistency with the meson-meson
scattering result would then require the exotic mesons to have 
suppressed coupling to the meson-meson channels, while they couple strongly
to baryon-antibaryon.  It was further shown in \cite{eight} that a
consistent description of the scattering of these exotic mesons with
the normal baryons implies the presence of exotic baryon states in the
$\overline{10},27$ or higher representations.  Again consistency with
the earlier result requires these exotic baryons to have suppressed
coupling to the normal meson-baryon channels, while they couple
strongly to baryon and exotic meson channels.

Fig. 1 shows a simple visual representation of these results in terms
of the quark duality diagrams \cite{nine}.  The first two diagrams
shows duality between the normal meson $(q\bar q)$ and baryon $(qqq)$
states in meson-meson and meson-baryon scattering.  The third diagram
shows that in baryon-antibaryon scattering the normal meson $(q\bar
q)$ Regge pole exchanges are dual to exotic meson $(qq \bar q\bar q)$
resonances and vice versa.  The last diagram shows duality between
normal meson $(q\bar q)$ and exotic baryon $(qqqq\bar q)$ states in the
scattering of exotic meson with normal baryon.  Fig. 2 shows the
coupling of the exotic meson $(qq\bar q\bar q)$ to (a) baryon-antibaryon 
and (b,c) meson-meson channels.  It is clear from this figure that
its couplings to the meson-meson channels can be suppressed by
requiring that (i) each pair of particles in the vertex is connected
by at least one quark line, and (ii) a quark-antiquark pair belonging
to the same particle can not annihilate one another \cite{ten}.  This
is a generalisation of the famous OZI rule \cite{eleven}, in which
case the forbidden coupling of $\phi(s\bar s)$ to $\rho(u\bar d)
\pi(d\bar u)$ had both these features, so that enforcing either one of
them was enough.  In view of the preferential coupling of the $(qq\bar
q\bar q)$ mesons to the baryon-antibaryon pair they were later
christened as Baryoniums in analogy with the heavy quarkonium states
\cite{twelve}.  One can easily check that the same rule would ensure
preferential coupling of exotic baryon $(qqqq\bar q)$ states to
Baryonium-baryon channels while suppressing it to normal meson-baryon
channels.  It is customary to call these exotic baryons simply
Pentaquark states.
\bigskip

\noindent{\Large\bf Evidences of Baryonium $(qq\bar q\bar q)$ Trajectory \\
Exchange:} 
\medskip

It is not widely known that duality and Regge phenomenology of
Baryonium states has had a number of spectacular successes.  There are
several strong experimental evidences suggesting the exchange of low
lying meson Regge trajectories, which contribute very significantly to
baryon-antibaryon scattering processes but not in meson-baryon or
meson-meson scattering.  They occur in two-body as well as inclusive
scattering processes and in both exotic and nonexotic channels.  These 
evidences came during the mid and late seventies, when the Regge
phenomenology had already gone out of fashion.  Therefore it will be
worthwhile to briefly summarise them here.

\begin{enumerate}
\item[{1)}] The first evidence came from the backward production of
normal $(q\bar q)$ meson resonances in the inclusive process
\cite{thirteen} 
\be
\pi^- p \delr \ p X_R (\pi^-,\rho^-,a^-_1,a^-_2,\rho^-_3),  
\label{two}
\ee
which proceeds via the exchange of $\Delta$ Regge trajectory.  Thus it
corresponds via optical theorem to the elastic scattering amplitude 
$A_{p\pi^- \bar p} \sim A_{p` \bar\Delta '}$, i.e. a baryon-antibaryon
amplitude.  As one sees from Fig. 3, the normal $(q\bar q)$ resonance
contributions to this channel are dual to Baryonium $(qq\bar q\bar q)$
trajectory exchange (3a) and vice versa (3b).  The Baryonium
trajectory was estimated from the duality prediction 
\be
\sigma_{\pi^-p \rightarrow pX_R} (M_R) \sim A_{p` \bar\Delta ' }
(M_R) \propto (M_R^2)^{\alpha_B(0)},
\label{three}
\ee
which indeed gave a low intercept $(\alpha_B (0) \sim -0.5)$.
Moreover these resonances were found to be dual to a normal meson
trajectory $(\alpha_m(0) \sim +0.5)$ in the forward production
process, which corresponds to a meson-meson channel \cite{fourteen}.
\item[{2)}] The difference of inclusive cross-sections,
\be
\Delta_{pp} (\pi^-) = \sigma_{\bar pp \rightarrow \pi^- X} -
\sigma_{pp \rightarrow \pi^- X} \sim A_{\bar p ` \Delta '} - A_{p `
\Delta '} ,
\label{four}
\ee
represents the difference between the corresponding antibaryon-baryon
and baryon-baryon scattering amplitudes.  Regge analysis of the energy
dependence of $\Delta_{pp} (\pi^-)$ showed the presence of a large
contribution from a low lying trajectory $(\alpha_B (0) \simeq -1)$
in addition to the normal meson trajectory $(\alpha_m \simeq 0.5)$
\cite{fifteen}.  In contrast the cross-section differences,
$\Delta_{Kp} (\pi^-)$ and $\Delta_{\pi p} (\pi^-)$, were well
described by the normal meson trajectory over the same energy range.

\item[{3)}] The energy dependence of the $\bar pp$ and $pp$ total
cross-section difference $(\Delta_{pp})$ shows a similar evidence
for a low lying Baryonium trajectory $(\alpha_B (0) \sim -0.5)$, in
addition to the normal meson trajectory $(\alpha_m (0) \sim +0.5)$,
while $\Delta_{Kp}$ is well fitted by the $\alpha_m (0)$ alone
\cite{sixteen}.  Although the size of the $\alpha_B (0)$ contribution
for $\Delta_{pp}$ is smaller than the inclusive case the precision of
the total cross-section data is much better, so that the signal is
quite unmistakable.

\item[{4)}] Finally the forward cross-sections for the inelastic
baryon-baryon scattering processes with exotic $t$-channel like $pn
\rightarrow \Delta^- \Delta^{++}$, $\bar pp \rightarrow \bar Y^{\star
+} Y^{\star -}$ and $\bar pp \rightarrow \bar\Sigma^+ \Sigma^-$ are at
least an order of magnitude higher than the corresponding meson-baryon
scattering processes \cite{sixteen,seventeen}.  This can be interpreted
as evidence for exotic Baryonium exchange contribution.  In particular
a fit to the $pn \rightarrow \Delta^- \Delta^{++}$ cross-section in
terms of a non-strange exotic baryonium exchange again gives a low
intercept of $\alpha_B (0) \sim -0.5$.  Moreover the coupling is
roughly of similar size as the Baryonium coupling to $\Delta_{pp}$
\cite{sixteen}.  

Thus the phenomenological evidences for Baryonium
trajectory exchange contributions are too strong to be fortuitous.  On
the other hand the phenomenological support for Baryonium resonance
spectroscopy is relatively modest, as we see below. 
\end{enumerate}
\newpage

\noindent {\Large\bf Status of Baryonium $(qq\bar q\bar q)$
Resonances:}
\medskip

Several resonances, both narrow and broad, have been reported to decay
mainly into the $\bar pp$ channel \cite{eighteen}.  But they are not
enough to reproduce the normal $(q\bar q)$ Regge trajectory exchange
contribution to this channel as per the duality diagram of Fig. 1c
\cite{nineteen}. So one would require many of the resonances below
the $\bar pp$ threshold to be $qq\bar q\bar q$ states.  In fact there
is a large excess of meson resonances in this region compared to the
potential model prediction for the $q\bar q$ states; and many of them
can be $qq\bar q\bar q$ states \cite{twenty}.  They have normal
widths, which may appear to conflict with the above mentioned OZI
rule.  But since they all have nonexotic quantum numbers, they could
have large mixing with the $q\bar q$ states.  This would account for
their normal decay widths into the meson-meson channels like those of
the glueball candidates.  But in any case it is outside the scope of
the duality and Regge formalism to predict the spectroscopy of
Baryonium resonances.  For this purpose several QCD inspired models
have been suggested -- e.g. the diquark models of \cite{twentyone} and
\cite{twentytwo}, and the colour junction model of \cite{twentythree}. 

Let us briefly discuss the models of \cite{twentyone} and
\cite{twentytwo}, since they are comparatively simple and closely related
to one another.  They envisage the $qq\bar q\bar q$ state to be
composed of a highly correlated diquark and antidiquark pair.  The
diquarks in the antisymmetric colour state $\overline{3}$ experience a
strongly attractive colour force.  Consequently there are diquark and
antidiquark $S$ wave bound states with the following colour $SU(3)$,
flavour $SU(3)$ and spin $SU(2)$ representations, which ensure their
overall antisymmetry -- i.e. 
\be
qq \rightarrow
(\overline{3}_c,\overline{3}_f,1),(\overline{3}_c,6_f,3); \ \bar q\bar q
\rightarrow (3_c,3_f,1), (3_c,\overline{6}_f,3).
\label{five}
\ee
The resulting colour singlet $qq\bar q\bar q$ states have the flavour
$SU(3)$ representations and spin,
\be
9_f, J = L; \ 18_f, J = L \pm 1; \ \overline{18}_f, J = L \pm 1; \
36_f, J = L \pm 2;
\label{six}
\ee
where $L$ is the orbital angular momentum between the diquark and the
antidiquark states.

The difference between the two models is that, unlike Chan and
Hogaasen \cite{twentytwo}, Jaffe \cite{twentyone} rejects OZI suppression
of the decay diagram of Fig. 2b.  Instead he treats it as a OZI
superallowed decay, which should have a larger width than the usual
OZI allowed decay of Fig. 2a.  Consequently he predicts the $qq\bar
q\bar q$ states to be even broader resonances than the $q\bar q$
mesons and decay mainly into the meson-meson channels.  But as
explained in \cite{eighteen,twentytwo} and the last paper of
ref. \cite{twentyone}, it will be more reasonable to 
associate the two types of decay with different states of the orbital
angular momentum $L$.  For $L=0$ the diquark and antidiquark states
would have overlapping wavefunctions.  This would facilitate
interaction between their consistuent quark and antiquark, leading to
their dissociation into two $q\bar q$ pairs (Fig. 2b).  But for $L
\geq 1$ the spatial separation between the diquark and antidiquark
would inhibit interaction between their constituents.  In this case it
will be easier for the colour $\overline{3}$ diquark to combine with a
colour $3$ quark from the vacuum to form a baryon (Fig. 2a).  Thus one
would expect the light $(L=0)$ $qq\bar q\bar q$ states to have
uninhibited decay into meson-meson channels, while the heavier ones
$(L \geq 1)$ decay mainly into baryon-antibaryon channels.

Both the models predict the lightest flavour nonet $(9_f)$ states of
eq. (\ref{six}) to occur below 1 GeV, while the lightest exotic states
($18_f$ and $36_f$) occur in the mass range of 1--2 GeV.  Indeed the
lightest $0^+$ states of the Particle Data Group \cite{twenty} have
been interpreted by Jaffe \cite{twentyone} as members of the lightest
flavour nonet state of $qq\bar q\bar q$ (see also \cite{twentyfour}).
However, there is no distinctive prediction of these models which can
be used as a smoking gun signal for the $qq\bar q\bar q$ resonance.
In fact all the $qq\bar q\bar q$ resonance candidates of
\cite{twenty} are also amenable to alternative interpretations.  This
is also true for the narrow charmed meson resonance, discovered
recently \cite{two}.  It goes without saying that an unmistakable
signal for a $qq\bar q\bar q$ meson will be a resonance seen in an
exotic channel.  But there is no credible evidence for such a signal
even after 35 years of their prediction.
\bigskip

\noindent {\Large\bf Status of the Exotic Baryon (Pentaquark)
Resonance:}
\medskip

This brings us finally to the exotic baryon (Pentaquark) state we had
started with.  As mentioned above the $\Theta^+$ (1540) resonance has
an exotic quantum number, $Y=2$, which means it belongs to
$\overline{10}$ or higher representation of flavour $SU(3)$ and has a
minimal quark content of $ud ud\bar s$.  Indeed this is the first
credible signal of an exotic hadron resonance, although it still needs
further confirmation.  Arguments for and against the resonance
interpretation of this peak are given e.g. in \cite{twentyfive} and
\cite{twentysix} respectively.  Meanwhile the NA49 collaboration
\cite{twentyseven} has presented evidence for a similar narrow peak in
another exotic baryon channel with $S = -2$ and $I = 3/2$, which could
be a member of the same $SU(3)$ multiplet as the $\Theta^+ (1540)$.
It should be mentioned 
here that the mass and the narrow width of $\Theta^+ (1540)$ were
predicted remarkably well by the Skyrme model along with a spin-parity
of ${1\over2}^+$ \cite{twentyeight}.  It has been emphasised by many authors
however that this agreement could be fortuitous (see
e.g. \cite{twentynine,thirty}).  Firstly this model approximates
QCD only in the $N_c \rightarrow \infty$ limit.  For $N_c = 3$, even
the two-flavour (chiral $SU(2) \times SU(2)$) model predicts spurious
exotic baryons with $(I,J) =
({5\over2},{5\over2}),({7\over2},{7\over2}),\cdots$, which have to be
discarded as artifacts. Moreover for $KN$ scattering it relies heavily
on chiral $SU(3) \times SU(3)$ symmetry, which is badly broken in
nature.  Therefore there are enough reasons for reserve.

Early quark models based on 5 uncorrelated quarks predicted a host
of light Pentaquark states with negative parity below 1.5 GeV
\cite{thirtyone}, while none have been found in the data.  Subsequent
studies have found evidence for strong diquark correlation even for
the normal $(qqq)$ baryon spectrum \cite{thirtytwo}.  For instance
the quark-diquark $(3_c - \overline{3}_c)$ model for baryons implies
similar colour dynamics as the $(q\bar q)$ mesons, which helps to
explain the equal slopes of the corresponding Regge trajectories.
Therefore the 
interpretations of the Pentaquark signal are mainly based on strongly
correlated diquarks \cite{thirty,thirtythree,thirtyfour,thirtyfive}.
In particular Jaffe and Wilczek \cite{thirty} have interpreted
$\Theta (ud ud\bar s)$ in terms of a simple extension of the diquark
model of \cite{twentyone}.  They assume each $ud$ pair to be the
lightest diquark of eq. (\ref{five}), i.e. a $S$ wave bound state,
antisymmetric in colour, flavour and spin
$(\overline{3}_c,\overline{3}_f,1)$.  The two $ud$ pairs must combine
into a colour antisymmetric $3_c$ in order to form a colour singlet
baryon with the $\bar s(\overline{3}_c)$.  Therefore these two diquark
$(ud)$ states must be spatially antisymmetric, which suggests $L=1$.
The $S$ wave bound state of $\bar s$ with this $(L=1)$ pair of scalar
diquarks will then have $J^P = {1\over2}^+$ and ${3\over2}^+$.  The
lighter state $({1\over2}^+)$ is identified with the $\Theta$ (1540).
Evidently the two $ud$ pairs form a flavour symmetric state
$\overline{6}_f$, which combines with the $\bar s (\overline{3}_f)$ to
form a $\overline{18}_f = \overline{10}_f + 8_f$. They predict the
mass splittings between the various baryon Isospin multiplets
belonging to this 18-plet representation, and compare it with the
corresponding spectrum of the chiral soliton (Skyrme) model
\cite{twentyeight}.  It is fair to say however that this model has no
natural explanation for the very narrow width of $\Theta$ (1540).
Moreover the results of \cite{thirty} have been strongly
criticised in \cite{thirtysix} using the effective Hamiltonian
approach to QCD, which gives much larger mass and width for $\Theta$
in this model.  Similarly, Shuryak and Zahed \cite{thirtythree} have
also estimated a larger $\Theta$ mass of $\sim 1880$ MeV in this model
due to the orbital excitation $(L=1)$ between the diquark pair.  So
they have suggested an alternative model, where the two diquarks are
in a relative $L=0$ state, but one of them is a tensor ($P$ wave)
diquark.  On the other hand an $I=2$ assignment for $\Theta$ has been
suggested in \cite{thirtyfour} to account for its narrow decay width
into the $KN$ channel via Isospin violation.  Other alternatives
have been suggested e.g. in ref. \cite{thirtyfive}.  Thus in short, the
jury is still out on the correct theoretical model for the $\Theta$
(1540) resonance.

Let me conclude by pointing out two model independent predictions for
this exotic baryon from the duality and Regge formalism. (i) The
exotic baryon $(qqqq\bar q)$ was predicted to have suppressed coupling
to the normal meson-baryon channel like $KN$, while having normal
coupling to the exotic meson $(qq\bar q\bar q)$ -- baryon $(qqq)$
channel \cite{eight}.  This feature can be justified in the
Jaffe-Wilczek model \cite{thirty} by a similar reasoning as that
used above to justify inhibited coupling of $L \geq 1$ Baryoniums to
meson-meson channels.  The spatial separation between the two $(ud)$
diquarks in relative $L=1$ state inhibits exchange of their constituents.
This suppresses the $\Theta (udud\bar s) \rightarrow K (d\bar s)
N(udu)$ decay, which in analogous to Fig. 2b.  But it allows $\Theta
(ud ud \bar s) \rightarrow \kappa (ud \bar s \bar u) N(udu)$, where one of
the $(ud)$ diquarks in $\overline{3}_c$ absorbs a $u$-quark from the vacuum
to form a colour singlet $N$, while the associated $\bar u$ is
absorbed by $\bar s$ to form a 4-quark meson (Baryonium) $\kappa$
(Fig. 1d).  Now the lightest 4-quark strange meson $\kappa$ is
predicted to be a scalar of 800--900 MeV mass in
\cite{twentyone,thirty}, while there is no clear experimental
candidate for such a state below 1 GeV.  So in either case the
$\Theta$ (1540) lies below the $\kappa N$ threshold and hence would be
expected to have only a narrow decay width into the $KN$ channel.  A
similar conclusion is reached in a recent study of $\Theta$ as a bound
state of $N$ and a scalar $\kappa$ of 800-900 MeV mass in the chiral
perturbation theory \cite{thirtyseven}.
(ii) Because of its uninhibited coupling to the $\kappa$ $(qq\bar
q\bar q)$ -- $N$ $(qqq)$ channel one expects significant $\Theta$
production cross-section in 
\be 
pn \rB \wedge \Theta,
\label{seven}
\ee
via the exchange of a $\kappa$ Baryonium $(qq\bar q\bar q)$ Regge trajectory
(Fig. 4).  As discussed above there are several experimental evidences
for a low lying Baryonium trajectory exchange.  Hence one expects a
significant cross-section for the above process only at relatively low
incident energy of $\lsim 4$ GeV, say.  Therefore this production
process can be looked for at any low energy high intensity proton
accelerator.  I hope the concerned experimental groups will undertake
this investigation. 

I thank Prof. V. Singh for drawing my attentionto the Pentaquark
papers and Prof. R.V. Gavai for discussions.

\newpage

\begin{figure}
\begin{center}
\hspace{10cm}
\epsfxsize=4in
\epsfbox{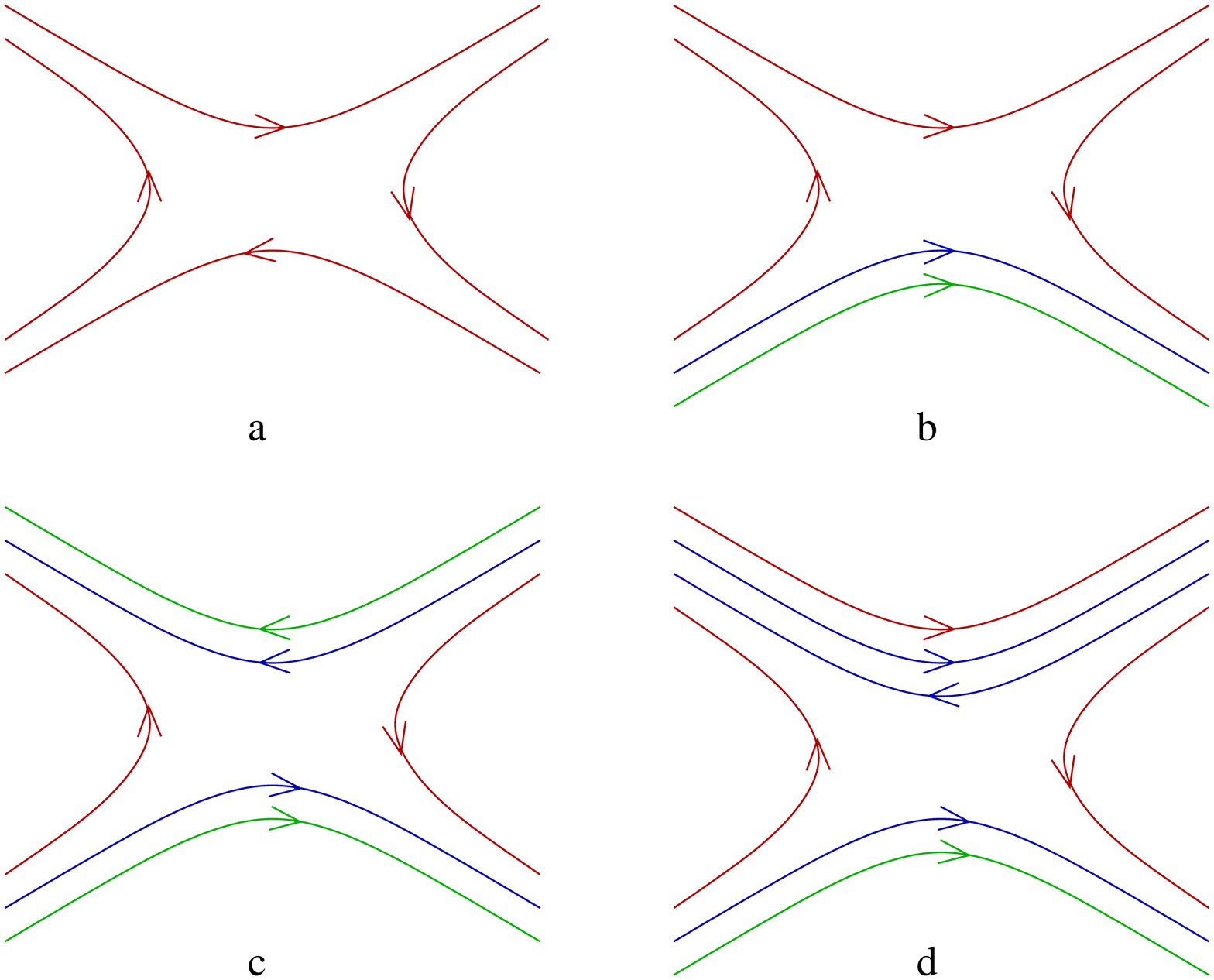}
\label{fig:0322fig1}
\bigskip
\caption{The quark duality diagrams for (a) meson-meson, (b)
meson-baryon, (c) baryon-antibaryon and (d) baryon-Baryonium
scattering.  The last one is the formation channel for Pentaquarks.}
\end{center}
\end{figure}
\vfill
\begin{figure}
\begin{center}
\hspace{10cm}
\epsfxsize=5in
\epsfbox{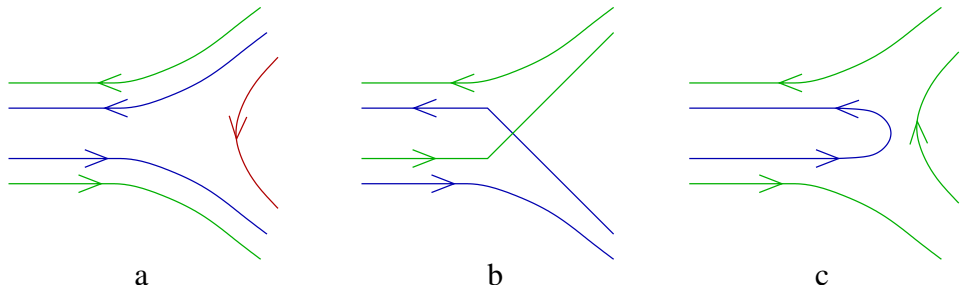}
\label{fig:0322fig2}
\bigskip\bigskip
\caption{The coupling of Baryonium into (a) baryon-antibaryon
and (b,c) meson-meson channels.  Only the 1st one is OZI allowed.}
\end{center}
\end{figure}

\newpage

\begin{figure}
\begin{center}
\hspace{10cm}
\epsfxsize=5in
\epsfbox{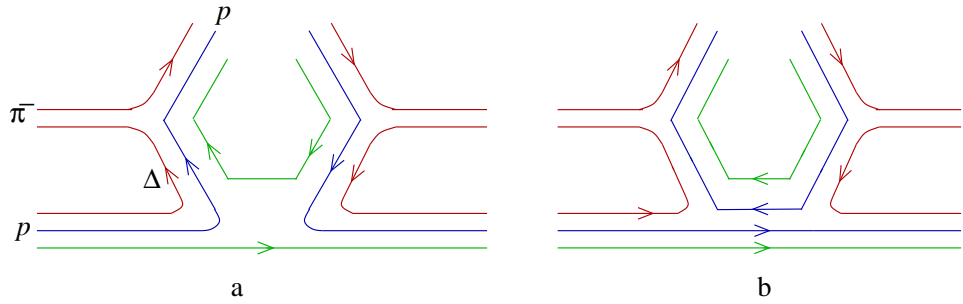}
\label{fig:0322fig4}
\bigskip\bigskip
\caption{The quark duality diagrams for backward production of (a)
normal meson and (b) Baryonium resonances.}
\end{center}
\end{figure}
\vfill
\begin{figure}
\begin{center}
\hspace{10cm}
\epsfxsize=5in
\epsfbox{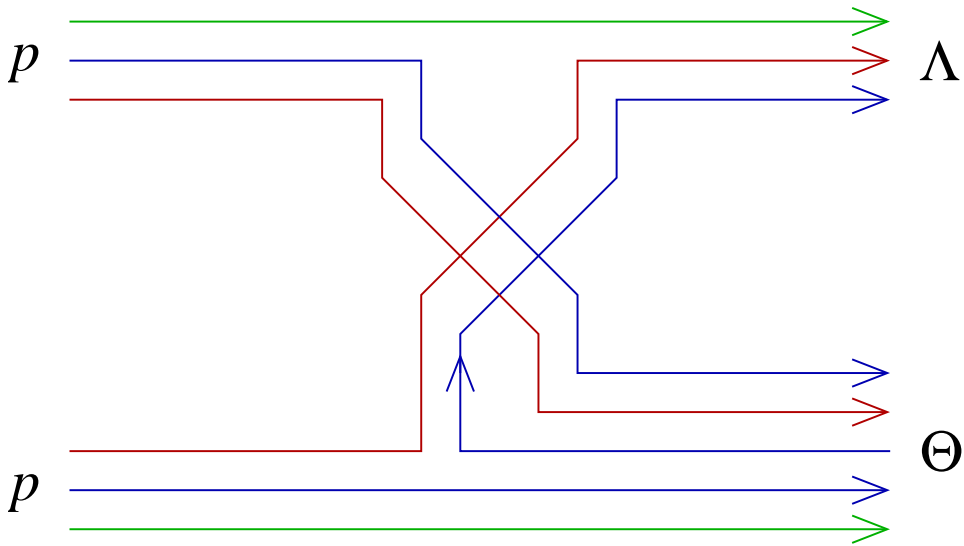}
\label{fig:0322fig3}
\bigskip\bigskip
\caption{The quark diagram for $\Theta$ production via Baryonium exchange.}
\end{center}
\end{figure}

\end{document}